\documentclass[cameraready]{Interspeech}

\title{TSExplorer: An interactive data annotation and exploration tool for time-series data}

\author[affiliation={1}, orcid=0000-0002-8714-6090, correspondingauthor]{Einari}{Vaaras}
\author[affiliation={2}, orcid=0000-0002-8031-2260]{Manu}{Airaksinen}
\author[affiliation={1}, orcid=0000-0002-0537-0946]{Okko}{Räsänen}

\address{
    $^1$ Signal Processing Research Centre, Tampere University, Finland \\
    $^2$ BABA Center, Department of Physiology, University of Helsinki, Finland
}

\email{einari.vaaras@tuni.fi, manu.airaksinen@helsinki.fi, okko.rasanen@tuni.fi}

\keywords{2D visualization, data annotation, human-computer interaction, data exploration, human-in-the-loop}

\usepackage{comment}

\begin{document}

\maketitle

\begin{abstract}
    We present TSExplorer, a cross-platform tool for interactive annotation and exploration of time-series data. The tool enables users to inspect high-dimensional datasets through multiple complementary 2D visualizations derived from high-dimensional feature representations. TSExplorer is designed as a general-purpose research tool supporting a wide range of workflows, including exploratory data analysis, annotation of unlabeled or partially-labeled datasets, comparison of feature representations, and post-hoc inspection and refinement of existing labels with interactive visual feedback.
\end{abstract}

\section{Introduction}

Many modern analysis pipelines rely on high-dimensional feature representations when working with time-series data, such as speech, video, physiological signals like electroencephalography, sensor-based recordings, or multimodal combinations of these modalities. These representations may be hand-crafted (e.g. log-mel spectrograms), or they can also be learned embeddings (e.g. wav2vec 2.0 \cite{wav2vec2}). While such representations can be powerful, common workflows for data annotation and analysis either discard them entirely or make limited use of their structure. In practice, data annotation is frequently performed in a sequential manner, and data analysis is typically based on static visualizations or aggregate summary statistics such as feature variances, class-wise averages, performance metrics, or correlation measures. Such static approaches make it difficult to explore individual samples in detail or to understand their relationships within the overall feature space. This limitation can be particularly noticeable for large-scale datasets, or when comparing how the dataset is structured with different feature representations.

To address these challenges, we present Time-Series Explorer (TSExplorer), a general-purpose graphical user interface (GUI) tool for time-series data annotation and exploration. TSExplorer allows users to freely explore high-dimensional datasets through interactive 2D visualizations (2DVs). Instead of following a single pre-defined workflow, TSExplorer is designed as a flexible tool that can be used for data annotation, analysis, and interpretation. Interactive 2D data visualization has previously been shown to support human-in-the-loop exploratory data analysis and interpretation (e.g. \cite{t_visne_article, clinical_data_2d_visualization_comparison}), as well as data annotation (e.g. \cite{umap_imaging_flow_cytometry, fallgren_interspeech}). TSExplorer is publicly available on GitHub\footnote{\url{https://github.com/SPEECHCOG/TSExplorer}}, and it has been previously described and evaluated in Vaaras et al. \cite{tsexplorer_journal_article}.

\begin{figure}[ht]
    \centering
    \includegraphics[width=0.47\textwidth]{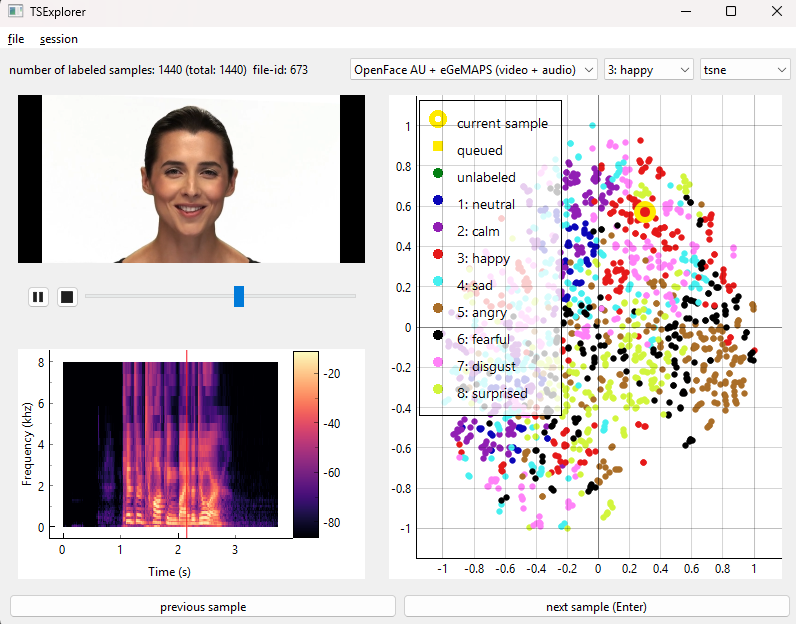}
    \caption{A screenshot of TSExplorer with the RAVDESS dataset~\cite{ravdess}. This example contains a video widget (top left), a spectrogram widget (bottom left), and a scatter widget (right).}
    \label{fig:tsexplorer_example_screenshot_ravdess}
\end{figure}

\section{Interactive Data Annotation and Exploration with TSExplorer}

The TSExplorer GUI (Figure \ref{fig:tsexplorer_example_screenshot_ravdess}) visualizes the entire dataset as a 2D scatter plot, where each data point corresponds to one data sample (e.g. an utterance in speech data). Selecting a point in the scatter plot presents corresponding views (e.g. audio, video, and signal waveforms) of the sample to the user. Users can alternate between t-distributed stochastic neighbor embedding (t-SNE) \cite{tsne}, principal component analysis (PCA), and uniform manifold approximation and projection (UMAP) \cite{umap_arxiv} 2DVs, each highlighting different aspects of the underlying high-dimensional data. Furthermore, users can switch between different high-dimensional feature representations, both within and across modalities, from which the 2DVs are computed.

\begin{figure*}[t]
    \centering
    \includegraphics[width=0.99\textwidth]{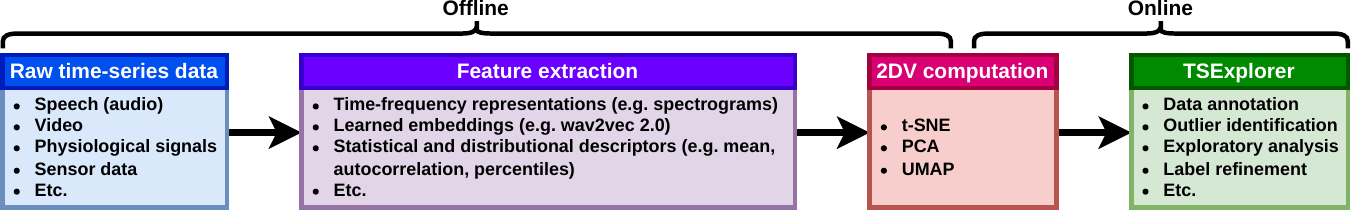}
    \caption{A block diagram of the TSExplorer workflow. First, high-dimensional features are computed offline from time-series data. Then, 2DVs are computed from these features, either pre-computed (offline) or within TSExplorer (online). Finally, the time-series data, high-dimensional features, and 2DVs are provided as inputs to TSExplorer for interactive data annotation and exploration.}
    \label{fig:tsexplorer_is_2026_show_and_tell_block_diagram}
\end{figure*}

In TSExplorer, users can select samples for inspection or annotation either by left-clicking data points in the scatter plot, or algorithmically using the \textit{next sample} button. Currently, there are three variants for algorithmic sample selection available: 1) \textit{Random}, where samples are selected at random; 2) \textit{Ordered}, where samples are selected sequentially from the dataset; 3) \textit{Farthest-first}, which selects samples using the farthest-first traversal algorithm \cite{faft_original}. Samples can also be enqueued for later inspection via right-clicking, after which they can be either sequentially browsed using the \textit{next sample} button, or by directly left-clicking them in the scatter plot. The scatter plot supports zooming using the scroll wheel of the mouse, and the data point sizes adapt dynamically. Labels can be modified at any given time either using a drop-down menu or with keyboard shortcuts. Sample colors in the scatter plot reflect their labels (green color for unlabeled samples by default), and the currently selected sample is highlighted with a large ring (yellow by default).

TSExplorer supports unlabeled, partially-labeled, and fully-labeled datasets. For unlabeled data, the tool can be used for data annotation and exploratory analysis, e.g. to help understand dataset structure or identify outliers. For partially-labeled data, TSExplorer supports incremental annotation with direct visual feedback on how labels distribute across the feature space, both within and across different feature representations. For fully-labeled data, the tool enables evaluating class separability or feature consistency, as well as label revision for data quality assurance or relabeling.

Currently, TSExplorer provides five widget types: audio, video, scatter, waveform, and spectrogram. The GUI layout is fully customizable, allowing users to modify widget size, location, and type, and to use multiple instances of the same widget. Each widget also contains configurable settings to support different use cases. For example, in the waveform widget, users can control waveform grouping, number of plots, spacing, colors, line thickness, axis labels, among other options as well. In addition to customizing existing widgets, TSExplorer is designed to be extensible: new widgets, 2DV methods, and algorithmic sample selection strategies can be added, and existing components can be modified in a straightforward manner.

\section{System Design and Implementation}

Figure \ref{fig:tsexplorer_is_2026_show_and_tell_block_diagram} illustrates the overall system structure. TSExplorer was designed with cross-platform compatibility and usage simplicity in mind, and it runs on Windows, Linux, and macOS systems. The tool is implemented in Python using PySide6 for the GUI and PyQtGraph for data visualization, and audio and video playback are handled using the VLC Media Player.

TSExplorer has modest computational and memory requirements, as feature representations are computed offline. During use, the backend primarily performs file input/output, with 2DV computation being the most demanding process. For example, computing a t-SNE embedding for 100,000 samples (160-dimensional) requires approximately 1.8 GB of RAM and 8 minutes on a single CPU core (3.9 GHz). To avoid repeated computation, TSExplorer saves the computed 2DVs, and pre-computing them is recommended for large datasets to reduce computational overhead. Memory usage scales with dataset size and feature configuration: For example, a speech dataset containing 100,000 samples (mean utterance duration 1.5 s) with pre-loaded audio and log-mel, MFCC, and F0 features and their corresponding 2DVs uses 5.1 GB of RAM. In contrast, having audio as individual files on disk and loading samples on demand reduces RAM usage below 0.4 GB. With no data loaded, TSExplorer idles at approximately 0.3 GB of RAM.

\section{Acknowledgments}

The authors are grateful to Santeri Heiskanen for developing the initial version of TSExplorer. This work was supported in part by the Research Council of Finland (grants no. 343498 and 371243) and in part by the Sigrid Jusélius Foundation. The author Einari Vaaras would also like to thank the Finnish Foundation for Technology Promotion for the encouragement grants and the Nokia Foundation for the Nokia Scholarship.

\section{Generative AI Use Disclosure}

Generative AI was used only in a limited role for programming support, such as syntax guidance and debugging.

\bibliographystyle{IEEEtran}
\bibliography{mybib}

\end{document}